\documentclass[a4paper,12pt]{article}
\usepackage{epsfig}
\usepackage[dvips,usenames]{color}
\usepackage{graphicx}
\usepackage{color}
\usepackage{colortbl}

\newlength{\dinwidth}
\newlength{\dinmargin}
\newcommand{\spur}[1]{\not\! #1 \,}

\begin{document}
\title{The rare decays $B^+_u\to\pi^+\ell^+\ell^-$, $\rho^+\ell^+\ell^-$ and $B^{0}_{d}\to
\ell^+\ell^-$ in the R-parity violating supersymmetry  }
\author{   Jian-Jun Wang$^{1,2}$, Ru-Min Wang$^{3}$, Yuan-Guo Xu$^1$, and
  Ya-Dong Yang$^1$\thanks{Corresponding author. E-mail address: yangyd@iopp.ccnu.edu.cn }  \\
{\small {$^1$ \it Institute of Particle Physics,
 Huazhong Normal University,  Wuhan, Hubei 430079, P.R.China }}
\\
{\small {$^2$ \it Department of Physics, Henan Normal University, XinXiang, Henan 453007, P.R.China}}
\\
{\small {$^3$ \it Department of Physics, Yonsei University, Seoul
120-479, Korea}}
 }
\maketitle

\begin{abstract}
We study the rare decays $B^{+}_u\to \pi^{+}\ell^{+}\ell^{-}$, $\rho^{+}\ell^{+}\ell^{-}$ and $B^{0}_d \to \ell^{+}
\ell^{-}(\ell=e,\mu)$ in the R-parity violating supersymmetric standard  model. From the latest  upper limits of
$\mathcal{B}(B^+_u\to\pi^+\ell^+\ell^-)$ and $\mathcal{B}(B^0_d\to\ell^+\ell^-)$, we have derived new upper
bounds on the relevant R-parity violating couplings products, which are stronger than the existing ones. Using the constrained parameter space,  we present the R-parity violating effects on the branching ratios  and the forward-backward asymmetries of these decays.  We find that $\mathcal{B}(B^0_d\to \ell^+\ell^-)$ and $\mathcal{B}(B^+_u\to\rho^+\ell^+\ell^-)$ could be enhanced several orders by the R-parity violating sneutrino  and   squark exchanges, respectively.   The R-parity violating effects on the dilepton invariant mass spectra of $B^+_u\to \pi^+\ell^+\ell^-$, $\rho^+\ell^+\ell^-$ and the normalized forward-backward asymmetry $\mathcal{A}_{FB}(B^+_u\to\pi^+\ell^+\ell^-)$ and $\mathcal{A}_{FB}(B^+_u\to\rho^+\ell^+\ell^-)$ are studied in detail.   Our results could be used to probe the R-parity violating effects and will correlate with  searches for the direct R-parity violating signals at the future experiments.
\end{abstract}
\vspace{0.5cm}
 \noindent {\bf PACS numbers: 13.20.He, 12.60.Jv, 12.15.Ji, 12.15.Mm}

\section{Introduction}

In the Standard Model (SM), rare $B$ decays  $B^+_u\to
\pi^+\ell^+\ell^-$, $\rho^+\ell^+\ell^-$ and $B^{0}_{d} \to
\ell^+\ell^-~(\ell=e,\mu)$ are  induced by $\bar{b}\to
\bar{d}\ell^{+}\ell^{-} $ flavor changing neutral current (FCNC) and
expected to be highly suppressed.  These decays  could be  an
important testing ground  of the SM and offer a complementary  means
to search for  new physics by probing the indirect effects of new
interactions.

The physical aspects of  these decays are  similar to the $b\to
s\ell^{+} \ell^{-} $ decays, which, due to its much larger rates,
have been studied much more extensively  in the SM \cite{eilam,
grinstein, Buras,   Jaus} and its various extensions
\cite{AliandHiller, aliball, Dai,  ltan, Lunghi, Xu}. Additionally
to the $b\to s\ell^{+} \ell^{-} $ decays,  the $b\to
d\ell^{+}\ell^{-} $ decays would serve  an independent test for new
flavor changing interactions. Experimental progresses  aimed to such
semileptonic FCNC decays have been made by \textit{BABAR}
\cite{babarbsll} and Belle \cite{bellebsll} with the measurements of
the forward-backward asymmetry ($\mathcal{A}_{FB}$) of  $B\to K^{*}
\ell^{+}\ell^{-}$ and the  branching ratios  of both $B\to K^{*}
\ell^{+}\ell^{-}$ and $K \ell^{+}\ell^{-}$. Furthermore,  the
radiative $b\to d \gamma$  penguin processes $B\to \rho \gamma$ and
$\omega \gamma$ have been measured by \textit{BABAR} \cite{babarbdg}
and Belle  \cite{bellebdg}. Along with these progresses,
\textit{BABAR},  Belle  and CDF have begun to probe $b\to
d\ell^{+}\ell^{-} $ decays. Recently,   \textit{BABAR}  has made a
search for  $B\to \pi\ell^+\ell^- $ and $B^{0}_d\to \ell^+\ell^-$
decays \cite{Aubert}. It's interesting  to note that the  new upper
limit of $\mathcal{B}(B\to \pi \ell^{+}\ell^{-})<9.1\times 10^{-8}$
at $90\%$ C.L. \cite{Aubert} has improved the previous upper limits
\cite{former} by four orders of magnitude. Meanwhile,  the  upper
limit of $\mathcal{B}(B^{0}_{d} \to \mu^{+}\mu^{-})$ has been
improved to be $\mathcal{B}(B^0_d\to \mu^{+}\mu^{-})<1.5 \times
10^{-8}$ at $90\%$ C.L. by CDF \cite{cdf}. These renewed upper
limits from \textit{BABAR} \cite{Aubert} and CDF \cite{cdf}  will
provide powerful constraints on new physics.

 In the literature,   many useful observables in $B\to \rho\ell^+\ell^-$ and
 $B\to \pi\ell^+\ell^-$ decays, such as branching ratio, $CP$-asymmetry,
 forward-backward asymmetry etc., have been investigated in the framework of the SM \cite{ kruger},
 the two Higgs doublet models \cite{Aliev2, itan2, Guray}   and supersymmetry models (SUSY)  \cite{voloshin}.
 With the new upper limits, it would be very worthy to  study these decays in other new physics models
 to derive bounds  on the relevant parameters.

 In this  paper,  we will study  these decays in the R-parity violating (RPV)
 supersymmetric standard model \cite{RPVm1, RPVm2, report}.
 The phenomenological constraints on the RPV  couplings have been studied extensively in  $B$ decays \cite{RPVstudy}.
 At first,  we  will update the SM expectations of these decays with the up-to-date inputs,
 such as  the new light-cone  QCD sum rules results for $B\to \pi(\rho)$ form factors  \cite{BallZwicky}
 and the electro-weak parameters \cite{PDG}.  To make realistic estimations, we have taken into account
 of the uncertainties of the input parameters by varying them randomly within $1\sigma$ variance.

The decays   $B^+_u\to \rho^+\ell^+\ell^-$, $B^+_u\to
\pi^+\ell^+\ell^-$ and $B^0_d\to\ell^+\ell^-$ are all induced at the
parton level by  the same set of the RPV coupling products in the
RPV SUSY.    From the latest experimental data and the theoretical
parameters, we will derive the new upper limits on the relevant RPV
coupling products. Then we will use the constrained parameter space
to predict the  RPV effects on the branching ratios of
 $B^+_u\to \rho^+\ell^+\ell^-$, $B^+_u\to
\pi^+\ell^+\ell^-$ and $B^0_d\to\ell^+\ell^-$ decays,  and the
forward-backward asymmetries of $B^+_u\to\rho^+\ell^+\ell^-$ and
$B^+_u\to\pi^+\ell^+\ell^-$. Moreover, we will compare the SM
predictions with the RPV predictions about the dilepton invariant
mass spectra and the normalized forward-backward asymmetries in
$B^+_u\to\rho^+\ell^+\ell^-$ and $B^+_u\to\pi^+\ell^+\ell^-$ decays.

The paper is organized as follows. In Sec. 2, we derive the
expressions for  $B^+_u\to \rho^+\ell^+\ell^-$, $B^+_u\to
\pi^+\ell^+\ell^-$ and $B^0_d\to\ell^+\ell^-$ processes in the RPV
SUSY. In Sec. 3,  our numerical analysis are presented. We display
the constrained parameter spaces which satisfy all the upper limits
of $\mathcal{B}(B^+_u\to\pi^+\ell^+\ell^-)$ and $\mathcal{B}(B^0_d\to
 \ell^+\ell^-)$, and then we use the constrained parameter spaces
 to present  the RPV effects on $\mathcal{A}_{FB}(B^+_u\to\rho^+\ell^+\ell^-)$,
 $\mathcal{A}_{FB}(B^+_u\to\pi^+\ell^+\ell^-)$,
$\mathcal{B}(B^+_u\to\pi^+\ell^+\ell^-)$,
$\mathcal{B}(B^+_u\to\rho^+\ell^+\ell^-)$ and $\mathcal{B}(B^0_d\to
\ell^+\ell^-)$. We also show
the RPV effects on the dilepton invariant mass spectra and the
normalized forward-backward asymmetries in
$B^+_u\to\rho^+\ell^+\ell^-$ and $B^+_u\to\pi^+\ell^+\ell^-$ decays.
Sec. 4 is devoted to our summary.

%%%%%%%%%%%%%%%%%%%%%%%%%%%%%%%%

\section{The theoretical frame for $B^+_u \to \pi^+\ell^+\ell^-$,  $B^+_u\to\rho^+\ell^+\ell^-$ and $B^0_d\to \ell^+\ell^- $}

\subsection{The decay branching ratios in the SM}

 \subsubsection{The semileptonic decays $B^{+}\to\pi^{+}\ell^{+}\ell^{-}$ and  $B^{+}\to\rho^{+}\ell^{+}\ell^{-}$ }

In the SM, the effective Hamiltonian for the $\bar{b}\to \bar{d}\ell^+\ell^-$
decay can be written as \cite{Buras, bdllkruger,  G.Buchalla}
\begin{eqnarray}
\mathcal{H}^{SM}_{eff}=-\frac{4G_F}{\sqrt{2}}V_{td}V^{*}_{tb}\left\{\sum^{10}_{i=1}C_{i}(\mu)\mathcal{O}_i(\mu)
+\lambda^{*}_u\sum^{2}_{i=1}C_{i}(\mu)[\mathcal{O}_{i}(\mu)-\mathcal{O}^{(u)}_i(\mu)]\right\},
\end{eqnarray}
where $\lambda^{*}_u\equiv\frac{V_{ub}^{*}V_{ud}}{V^{*}_{tb}V_{td}}$. The
explicit form of all operators $\mathcal{O}_i(\mu)$  and  the Wilson
coefficients $C_i(\mu)$ calculated in the naive dimensional
regularization (NDR) scheme can be found in  \cite{Buras}. The
effective Hamiltonian leads the QCD corrected matrix  element for
$\bar{b}\to \bar{d}\ell^+\ell^-$ \cite{ Buras,bdllkruger}
\begin{eqnarray}
\mathcal{M}^{SM}(\bar{b}\to \bar{d}
\ell^+\ell^-)&=&\frac{G_F\alpha_{e}}{\sqrt{2}\pi}V^{*}_{tb}V_{td}\Biggl\{C^{eff}_9
(\bar{b}\gamma_\mu P_{L} d)(\bar{\ell}\gamma^\mu
\ell)+C_{10}(\bar{b}\gamma_\mu
P_{L} d)(\bar{\ell}\gamma^\mu\gamma_5\ell)\nonumber\\
&&-2\hat{m}_{b}
C^{eff}_7\left(\bar{b}i\sigma_{\mu\nu}\frac{\hat{q}^\nu}{\hat{s}}P_{L} d\right)
(\bar{\ell}\gamma^\mu \ell)\Biggl\},\label{quarkM}
\end{eqnarray}
with $ P_{L,R} \equiv (1\mp\gamma_5)/2, s = q^2$ and $q =
p_{+} + p_{-}$ ( $p_{\pm}$ are the four-momenta of the leptons). We take
$m_d/m_b=0$, but keep the leptons mass. The hat denotes
normalization in terms of the B-meson mass, $m_B$, e.g.
$\hat{s}=s/m_B^2$, $\hat{m}_q=m_q/m_B$.

$C^{eff}_{9}(\mu)$ contains the one-gluon corrections to
$\mathcal{O}_{9}$ ($\omega(\hat{s})$ term)  \cite{kuhn}  and the
corrections of  four-quark operators $\mathcal{O}_{1-6}$ and
$\mathcal{O}^{u}_{1,2}$ in Eq.(1)  \cite{misiak},    which  can be
written as \cite{bdllkruger}
\begin{eqnarray}
C^{eff}_9(\mu)=\xi_1+\lambda^{*}_u\xi_2,
\end{eqnarray}
with
\begin{eqnarray}
\xi_1&=&\hat{C}^{NDR}_9\left[1+\frac{\alpha_s(\mu)}{\pi}\omega(\hat{s})\right]+g(\hat{m}_c,\hat{s})
\Big(3C_1+C_2+3C_3+C_4+3C_5+C_6\Big)\nonumber\\
&&-\frac{1}{2}g(1,\hat{s})\Big(4C_3+4C_4+3C_5+C_6\Big)
-\frac{1}{2}g(0,\hat{s})\Big(C_3+3C_4\Big)\nonumber\\
&&+\frac{2}{9}\Big(3C_3+C_4+3C_5+C_6\Big), \\
\xi_2 &=& \Big[g(\hat{m}_c,
\hat{s})-g(0,\hat{s})\Big]\Big(3C_1+C_2\Big).
\end{eqnarray}
In addition to the short distance contributions, $\bar{b}\to
\bar{d}\ell^+\ell^-$ decay also receives long distance contributions
from both on- and off-shell
 vector mesons, which can be incorporated in $C^{eff}_9$ by the replacement  \cite{VMD1,VMD2}
 \begin{eqnarray}
 g(\hat{m}_c, \hat{s}) \to    g(\hat{m}_c, \hat{s}) -  \frac{3\pi}{\alpha^{2}_{e}}  \sum_{V_i=\psi, \psi' ...} \kappa
\frac{m_{V_{i}}\Gamma(V_i \to\ell^+\ell^- )}{ s- m^{2}_{V_{i}} +i
\Gamma_{V_{i}} m_{V_{i}}}.
  \end{eqnarray}
This issue has been extensively discussed in literature
\cite{Babu,bdll}.

With these formulas, one can derive the decay amplitudes for $B^{+}\to\pi^{+}\ell^{+}\ell^{-}$ 
and  $\rho^{+}\ell^{+}\ell^{-}$ decays. Recently  the form factors for  $B\to\pi $ and $B\to \rho$
transitions have been improved in the framework of light-cone QCD
sum rules  \cite{BallZwicky}, with  one-loop radiative corrections
to twist-2 and twist-3 contributions, and leading order twist-4
corrections.   Using  the form-factors defined in  \cite{BallZwicky}
and the matrix in Eq.(\ref{quarkM}),  we get the following
amplitudes for  $B^+_u\to\pi^+\ell^+\ell^-$ and
$B^+_u\to\rho^+\ell^+\ell^-$ decays:
\begin{eqnarray}
\mathcal{M}^{SM}(B\to M \ell^+ \ell^-
)=\frac{G_F\alpha_{e}}{2\sqrt{2}~\pi}
V_{td}V^{*}_{tb}m_B\left[\mathcal{T}_{1\mu}(\bar{\ell}\gamma^\mu
\ell)+\mathcal{T}_{2\mu}(\bar{\ell}\gamma^\mu
\gamma_5\ell)\right],
\end{eqnarray}
where for $B^+_u\to\pi^+\ell^+\ell^-$,
\begin{eqnarray}
\mathcal{T}_{1\mu}=A'(\hat{s})\hat{p}_\mu+B'(\hat{s})\hat{q}_\mu,\\
\mathcal{T}_{2\mu}=C'(\hat{s})\hat{p}_\mu+D'(\hat{s})\hat{q}_\mu,
\end{eqnarray}
 and for $B^+_u\to
\rho^+\ell^+\ell^-$,
\begin{eqnarray}
\mathcal{T}_{1\mu}&=&A(\hat{s})\epsilon_{\mu\rho\alpha\beta}\epsilon^{*\rho}
\hat{p}^\alpha_B\hat{p}^\beta_{\rho}-iB(\hat{s})\epsilon^*_\mu
+iC(\hat{s})(\epsilon^*\cdot\hat{p}_B)\hat{p}_\mu+iD(\hat{s})(\epsilon^*\cdot\hat{p}_B)\hat{q}_\mu,\\
\mathcal{T}_{2\mu}&=&E(\hat{s})\epsilon_{\mu\rho\alpha\beta}
\epsilon^{*\rho}\hat{p}^\alpha_B\hat{p}^\beta_{\rho}-iF(\hat{s})\epsilon^*_\mu
+iG(\hat{s})(\epsilon^*\cdot\hat{p}_B)\hat{p}_\mu+iH(\hat{s})(\epsilon^*\cdot\hat{p}_B)\hat{q}_\mu,
\end{eqnarray}
with $ p\equiv p_B+p_M$($M= \pi^+$ or $\rho^+$ ).

The auxiliary functions in  $\mathcal{T}_{1,2 \mu}$ are defined as
\cite{aliball}
%%%
\begin{eqnarray}
A'(\hat{s})&=&C^{eff}_9(\hat{s})f_+(\hat{s})
+\frac{2\hat{m}_b}{1+\hat{m}_\pi}C^{eff}_7f_T(\hat{s}),\\
B'(\hat{s})&=&C^{eff}_9(\hat{s})f_-(\hat{s})
-\frac{2\hat{m}_b}{\hat{s}}(1-\hat{m}_\pi)C^{eff}_7f_T(\hat{s}),\\
C'(\hat{s})&=&C_{10}f_+(\hat{s}),\\
D'(\hat{s})&=&C_{10}f_-(\hat{s}),\\
A(\hat{s})&=&\frac{2}{1+\hat{m}_\rho}C^{eff}_9(\hat{s})
V(\hat{s})+\frac{4\hat{m}_b}{\hat{s}}C^{eff}_7T_1(\hat{s}),\\
B(\hat{s})&=&(1+\hat{m}_\rho)\left[C^{eff}_9(\hat{s})A_1(\hat{s})
+\frac{2\hat{m}_b}{\hat{s}}(1-\hat{m}_\rho)C^{eff}_7T_2(\hat{s})\right],\\
C(\hat{s})&=&\frac{1}{1-\hat{m}^2_\rho}\left[(1-\hat{m}_\rho)C^{eff}_9(\hat{s})A_2(\hat{s})
+2\hat{m}_bC^{eff}_7\left(T_3(\hat{s})
+\frac{1-\hat{m}_\rho}{\hat{s}}T_2(\hat{s})\right)\right],\\
D(\hat{s})&=&\frac{1}{\hat{s}}\Big[C^{eff}_{9}(\hat{s})
\Big((1+\hat{m}_\rho)A_1(\hat{s})-(1-\hat{m}_\rho)A_2(\hat{s})
-2\hat{m}_\rho A_0(\hat{s})\Big)\nonumber\\&&-2\hat{m}_bC^{eff}_7T_3(\hat{s})\Big],\\
E(\hat{s})&=&\frac{2}{1+\hat{m}_\rho}C_{10}V(\hat{s}),\\
F(\hat{s})&=&(1+\hat{m}_\rho)C_{10}A_1(\hat{s}),\\
G(\hat{s})&=&\frac{1}{1+\hat{m}_\rho}C_{10}A_2(\hat{s}),\\
H(\hat{s})&=&\frac{1}{\hat{s}}C_{10}\Big[(1+\hat{m}_\rho)A_1(\hat{s})
-(1-\hat{m}_\rho)A_2(\hat{s})-2\hat{m}_\rho A_0(\hat{s})\Big].
\end{eqnarray}

 The kinematic variables ($\hat{s}$, $\hat{u}$)  are chosen to be
\begin{eqnarray}
\hat{s}&=&\hat{q}^2=(\hat{p}_++\hat{p}_-)^2,\\
\hat{u}&=&(\hat{p}_B-\hat{p}_-)^2-(\hat{p}_B-\hat{p}_+)^2,
\end{eqnarray}
which are bounded as
\begin{eqnarray}
(2\hat{m}_\ell)^2\leq&\hat{s}&\leq(1-\hat{m}_{\rho})^2,\\
-\hat{u}(\hat{s})\leq&\hat{u}&\leq\hat{u}(\hat{s}),
\end{eqnarray}
with $\hat{m}_\ell=m_{\ell}/m_B$ and
\begin{eqnarray}
\hat{u}(\hat{s})&=&\sqrt{\lambda\big(1-4\frac{\hat{m}^2_{\ell}}{\hat{s}}\big)},\\
\lambda&\equiv &\lambda(1,\hat{m}^2_{\pi,\rho},\hat{s})\nonumber\\
&=&1+\hat{m}^4_{\pi,\rho}
+\hat{s}^2-2\hat{s}-2\hat{m}^2_{\pi,\rho}(1+\hat{s}).
\end{eqnarray}
Note that the variable $\hat{u}$ corresponds to $\theta$, the angle
between the three-momentum of the $\ell^+$ lepton and the $B$ meson
in the dilepton center-of-mass system (CMS) frame, through the
relation $\hat{u} =-\hat{u}(s)\mbox{cos}\theta $ \cite{att1991}.
Keeping the lepton mass,  the double differential decay branching
ratios $\mathcal{B}^{\pi^+}$ and $\mathcal{B}^{\rho^+}$ for the
decays $B^+_u\to \pi^+\ell^+\ell^-$ and $B^+_u\to
\rho^+\ell^+\ell^-$, respectively,  are found to be
%%%%
%%%%
\begin{eqnarray}
\frac{d^2\mathcal{B}^{\pi^+}_{SM}}{d\hat{s}d\hat{u}}&=&\tau_B
\frac{G^2_F\alpha_{e}^2m_B^5}{2^{11}\pi^5}|V_{td}V^*_{tb}|^2
 \Bigg\{(|A'|^2+|C'|^2)(\lambda-\hat{u}^2)\nonumber\\
&&+|C'|^24\hat{m}^2_\ell(2+2\hat{m}^2_{\pi^+}-\hat{s})
+Re(C'D'^*)8\hat{m}^2_\ell(1-\hat{m}^2_{\pi^+})+|D'|^24\hat{m}^2_\ell\hat{s}
\Bigg\},  \label{Bpi} \\
%%%
\frac{d^2\mathcal{B}^{\rho^+}_{SM}}{d\hat{s}d\hat{u}}
&=&\tau_B\frac{G^2_F\alpha_{e}^2m_B^5}{2^{11}\pi^5}|V_{td}V^*_{tb}|^2\nonumber\\
&&\times\left\{\frac{|A|^2}{4}\Big(\hat{s}(\lambda+\hat{u}^2)+4\hat{m}^2_\ell\lambda\Big)
+\frac{|E|^2}{4}\Big(\hat{s}(\lambda+\hat{u}^2)
-4\hat{m}^2_\ell\lambda\Big)\right.\nonumber\\
&&+\frac{1}{4\hat{m}^2_{\rho^+}}\Big[|B|^2\Big(\lambda-\hat{u}^2
+8\hat{m}^2_{\rho^+}(\hat{s}+2\hat{m}^2_\ell)\Big)
+|F|^2\Big(\lambda-\hat{u}^2+8\hat{m}^2_{\rho^+}(\hat{s}-4\hat{m}^2_\ell)\Big)\Big]\nonumber\\
&&-2\hat{s}\hat{u}\Big[Re(BE^*)+Re(AF^*)\Big]\nonumber\\
&&+\frac{\lambda}{4\hat{m}^2_{\rho^+}}\Big[|C|^2(\lambda-\hat{u}^2)+
|G|^2(\lambda-\hat{u}^2+4\hat{m}^2_\ell(2+2\hat{m}^2_{\rho^+}-\hat{s})\Big)\Big]\nonumber\\
&&-\frac{1}{2\hat{m}^2_{\rho^+}}\Big[Re(BC^*)(1-\hat{m}^2_{\rho^+}-\hat{s})(\lambda-\hat{u}^2)\nonumber\\
&&~~~~~~~~~+Re(FG^*)\Big((1-\hat{m}^2_{\rho^+}
-\hat{s})(\lambda-\hat{u}^2)+4\hat{m}^2_\ell\lambda\Big)\Big]\nonumber\\
&&\left.-2\frac{\hat{m}^2_\ell}{\hat{m}^2_{\rho^+}}
\lambda\Big[Re(FH^*)-Re(GH^*)(1-\hat{m}^2_{\rho^+})\Big]
+|H|^2\frac{\hat{m}^2_\ell}{\hat{m}^2_{\rho^+}}\hat{s}\lambda
\right\}.
\end{eqnarray}
%%%%%%%%

\subsubsection{The pure leptonic decays $B_{d}^{0} \to \ell^+\ell^-$ }

In the SM,  the purely leptonic decays  $B_{d}^{0}\to \ell^+\ell^-$
proceed trough electroweak penguin diagrams with Z exchange as well
as W box diagrams. The photon penguin contribution is forbidden by
the conservation of electromagnetic current.   The effective
Hamiltonian  for these decays is  \cite{buchalla93}
\begin{eqnarray}
 \mathcal{H}^{SM}_{eff}(B^{0}_{d}\to \ell^+\ell^-)
 =-\frac{G_{F}}{\sqrt{2}}\frac{\alpha_{e}}{2 \pi \mbox{sin}^{2}\theta_W}
 V_{td}V^{*}_{tb}Y(x_t)(\overline{b}d)_{V-A}(\overline{\ell}\ell)_{V-A}, \label{pureH}
\end{eqnarray}
 where $x_{t}\equiv m_{t}^{2}/m_W^2$, which leads to  the  branching ratio
\begin{eqnarray}
\mathcal{B}^{SM}(B_d\to \ell^+\ell^-)&=&\frac{\tau_{B_d}}{16\pi
m_{B_d}}\sqrt{1-4\hat{m}^2_\ell}\left|\mathcal{M}^{SM}(B_d\to
\ell^+\ell^-)\right|^2\nonumber\\
&=&\tau_{B_d}\frac{G^2_F}{\pi}\left(\frac{\alpha_{e}}{4\pi
\mbox{sin}^2\theta_W}\right)^2f^2_{B_d}m^2_{\ell}m_{B_d}
\sqrt{1-4\hat{m}^2_{\ell}}\left|V_{td}V^*_{tb}\right|^2 |Y(x_t)|^{2}.
\label{pureB}
\end{eqnarray}
In the SM, as shown by Eq.(\ref{pureB}),  the decays are suppressed
by helicity conservation (the factor $m^{2}_{\ell}/m^{2}_{B}$ ) in
addition to the  small $V_{td}$. However,  they are very sensitive
to new physics with pseudo-scalar interactions.

\subsection{The decay amplitudes in the RPV SUSY}

In the most general superpotential of the minimal supersymmetric
standard model, the RPV superpotential is given by \cite{RPVm1,
RPVm2, report}
\begin{eqnarray}
\mathcal{W}_{\spur{R_p}}&=&\mu_i\hat{L}_i\hat{H}_u+\frac{1}{2}
\lambda_{[ij]k}\hat{L}_i\hat{L}_j\hat{E}^c_k+
\lambda'_{ijk}\hat{L}_i\hat{Q}_j\hat{D}^c_k+\frac{1}{2}
\lambda''_{i[jk]}\hat{U}^c_i\hat{D}^c_j\hat{D}^c_k, \label{rpv}
\end{eqnarray}
where $i$, $j$, $k$ = 1, 2, 3 are generation indices and $c$ denotes
a charge conjugate field. $\hat{L}_i(\hat{Q}_i)$ are the
lepton(quark) $SU(2)_L$ doublet superfields, $\hat{E}^c$,
$\hat{U}^c$ and $\hat{D}^c$ are the singlet superfields,
respectively. The bilinear RPV superpotential terms
$\mu_i\hat{L}_i\hat{H}_u$ can be rotated away by suitable redefining
the lepton and Higgs superfields  \cite{report}. However, the
rotation will generate a soft SUSY breaking bilinear term which
would affect our calculation through penguin level. The
processes discussed in this paper could be induced by the tree-level
RPV couplings, so that we would neglect sub-leading RPV penguin
contributions in this study. The $\lambda$ and $\lambda'$ couplings
in Eq.(\ref{rpv}) break the lepton number, while the $\lambda''$
couplings break the baryon number. There are 27 $\lambda'_{ijk}$
couplings, 9 $\lambda_{ijk}$ and 9 $\lambda''_{ijk}$ couplings.
$\lambda_{[ij]k}$ are antisymmetric with respect to their first two
indices, and $\lambda''_{i[jk]}$ are antisymmetric with $j$ and $k$.
\begin{figure}[h]
\begin{center}
\begin{tabular}{c}
\includegraphics[scale=1]{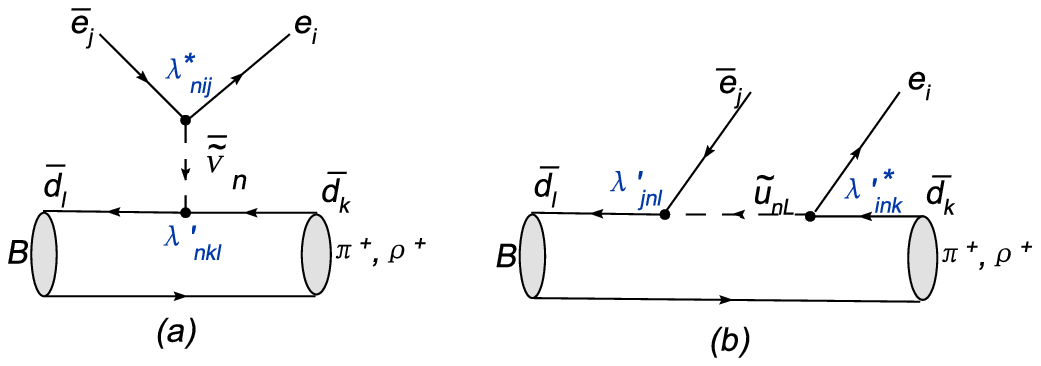}
\end{tabular}
\end{center}
\vspace{-0.8cm} \caption{ The RPV contributions to $B^+_u\to
\pi^+(\rho^+)\ell^+\ell^-$ due to the sneutrino  and squark
exchange.} \label{semifig}
\begin{center}
\begin{tabular}{c}
\includegraphics[scale=1]{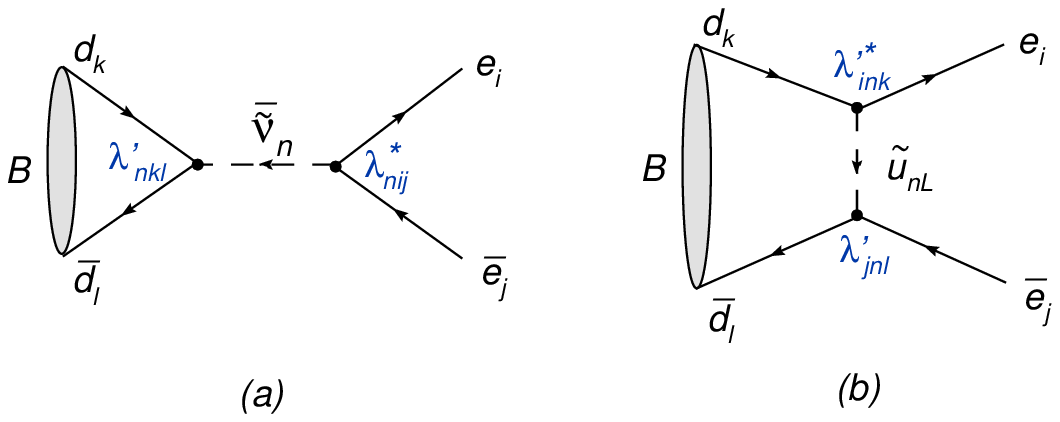}
\end{tabular}
\end{center}
\vspace{-0.8cm} \caption{ The RPV contributions to $B^0_d\to
\ell^+\ell^-$ due to the sneutrino  and squark exchange.}
\label{purefig}
\end{figure}
All the processes considered in this paper involve terms in the RPV
Hamiltonian with two leptons and two quarks as external fields.
From Eq.(\ref{rpv}), the Hamiltonian of $\bar{b}\to \bar{d}\ell^{+}\ell^{-}$  process
due to the squarks and sneutrinos exchange  are
\begin{eqnarray}
 \mathcal{H}^{\spur{R_p}}_{eff}&=&-\frac{1}{2}\sum_i \frac{\lambda'_{jik}\lambda'^{*}_{lin}}
 {m^2_{\tilde{u}_{iL}}}(\bar{d}_k\gamma^\mu P_R d_n)(\bar{\ell}_l\gamma_\mu
P_L\ell_j) \nonumber\\
 && +\sum_i\left\{\frac{\lambda_{ijk}\lambda'^{*}_{imn}}{m^2_{\tilde{\nu}_{iL}}}
 (\bar{d}_mP_Rd_n)(\bar{\ell}_kP_L\ell_j)+
\frac{\lambda^*_{ijk}\lambda'_{imn}}{m^2_{\tilde{\nu}_{iL}}}
(\bar{d}_nP_Ld_m)(\bar{\ell}_jP_R\ell_k)\right\}. \label{RPVH}
\end{eqnarray}
The RPV Feynman diagrams of $B^+_u\to \pi^+\ell^+\ell^-$, $B^+_u\to
\rho^+\ell^+\ell^-$ and $B^0_d\to \ell^+\ell^-$ are displayed in
Fig.\ref{semifig} and Fig.\ref{purefig}, respectively.

From Eq.(\ref{RPVH}), we can obtain the RPV decay amplitude for
$B^+_u\to \pi^+\ell^+\ell^-$
\begin{eqnarray}
\mathcal{M}^{\spur{R_p}}(B^+_u\to
\pi^+\ell^+\ell^-)=\Upsilon_{\tilde{u}}\Big(\bar{\ell}_k(\spur{p}_B+\spur{p}_{\pi^+})(1-\gamma_5)\ell_j\Big)
+\Upsilon_{\tilde{\nu}}\Big(\bar{\ell}_k(1-\gamma_5)\ell_j\Big)
+\Upsilon'_{\tilde{\nu}}\Big(\bar{\ell}_k(1+\gamma_5)\ell_j\Big),
\end{eqnarray}
where
\begin{eqnarray}
&&\Upsilon_{\tilde{u}}=-\sum_i\frac{\lambda'_{ji3}\lambda_{ki1}'^{*}}{8m^2_{\tilde{u}_{iL}}}
f_+^{B^+_u\to \pi^+}(\hat{s}),\\
&&\Upsilon_{\tilde{\nu}}=\sum_i\frac{\lambda_{ijk}\lambda_{i31}'^{*}}{4m^2_{\tilde{\nu}_{iL}}}
f_+^{B^+_u\to \pi^+}(\hat{s})\frac{m^2_B-m^2_{\pi^+}}{\overline{m}_b-\overline{m}_d},\\
&&\Upsilon'_{\tilde{\nu}}=\sum_i\frac{\lambda^*_{ikj}\lambda'_{i13}}{4m^2_{\tilde{\nu}_{iL}}}
f_+^{B^+_u\to
\pi^+}(\hat{s})\frac{m^2_B-m^2_{\pi^+}}{\overline{m}_b-\overline{m}_d},
\end{eqnarray}
 with $\overline{m}_b$ and $\overline{m}_d$ are the quark's running
masses at the scale $m_b$.

  For $B^+_u\to \rho^+\ell^+\ell^-$, the RPV amplitude is
\begin{eqnarray}
\mathcal{M}^{\spur{R_p}}(B^+_u\to
\rho^+\ell^+\ell^-)=\mathcal{T}_{3\mu}\Big(\bar{\ell}_k\gamma^\mu(1-\gamma_5)\ell_j\Big)
+~\Phi_{\tilde{\nu}}\Big(\bar{\ell}_k(1-\gamma_5)\ell_j\Big)
+~\Phi'_{\tilde{\nu}}\Big(\bar{\ell}_k(1+\gamma_5)\ell_j\Big),
\end{eqnarray}
where
\begin{eqnarray}
\mathcal{T}_{3\mu}&=&I(\hat{s})\epsilon_{\mu\rho\alpha\beta}\epsilon^{*\rho}
\hat{p}^\alpha_B\hat{p}^\beta_{\rho^+}-iJ(\hat{s})\epsilon^*_\mu
+iK(\hat{s})(\epsilon^*\cdot\hat{p}_B)\hat{p}_\mu+iL(\hat{s})(\epsilon^*\cdot\hat{p}_B)\hat{q}_\mu,\\
\Phi_{\tilde{\nu}}&=&\sum_i\frac{\lambda_{ijk}\lambda_{i31}'^{*}}{4m^2_{\tilde{\nu}_{iL}}}
\left[-\frac{i}{2}\frac{A_0^{B^+_u\to
\rho^+}(\hat{s})}{\overline{m}_b+\overline{m}_d}
\lambda^{\frac{1}{2}}m^2_B\right],\\
\Phi'_{\tilde{\nu}}&=&\sum_i\frac{\lambda^*_{ikj}\lambda'_{i13}}{4m^2_{\tilde{\nu}_{iL}}}
\left[\frac{i}{2}\frac{A_0^{B^+_u\to
\rho^+}(\hat{s})}{\overline{m}_b+\overline{m}_d}\lambda^{\frac{1}{2}}m^2_B\right],
\end{eqnarray}
and the auxiliary functions above are found to be
\begin{eqnarray}
I(\hat{s})&=&-\sum_i\frac{\lambda'_{ji3}\lambda_{ki1}'^{*}}{8m^2_{\tilde{u}_{iL}}}
\left[\frac{2V^{B^+_u\to \rho^+}(\hat{s})}{m_B+m_{\rho^+}}m^2_B\right],\\
J(\hat{s})&=&\sum_i\frac{\lambda'_{ji3}\lambda_{ki1}'^{*}}{8m^2_{\tilde{u}_{iL}}}
\left[(m_B+m_{\rho^+})A_1^{B^+_u\to \rho^+}(\hat{s})\right],\\
K(\hat{s})&=&-\sum_i\frac{\lambda'_{ji3}\lambda_{ki1}'^{*}}{8m^2_{\tilde{u}_{iL}}}
\left[\frac{A_2^{B^+_u\to \rho^+}(\hat{s})}{m_B+m_{\rho^+}}m_B^2\right],\\
L(\hat{s})&=&-\sum_i\frac{\lambda'_{ji3}\lambda_{ki1}'^{*}}{8m^2_{\tilde{u}_{iL}}}
\left[\frac{~2m_{\rho^+}}{\hat{s}}\Big(A_3^{B^+_u\to
\rho^+}(\hat{s})-A_0^{B^+_u\to \rho^+}(\hat{s})\Big)\right].
\end{eqnarray}

For the pure leptonic decays $B^0_d\to \ell^+\ell^-$,  the RPV
amplitude is
\begin{eqnarray}
\mathcal{M}^{\spur{R_p}}(B_d\to
\ell^+\ell^-)=\Omega_{\tilde{u}}\Big(\bar{\ell}_k\spur{p}_B(1-\gamma_5)\ell_j\Big)
+\Omega_{\tilde{\nu}}\Big(\bar{\ell}_k(1-\gamma_5)\ell_j\Big)
+\Omega'_{\tilde{\nu}}\Big(\bar{\ell}_k(1+\gamma_5)\ell_j\Big),
\end{eqnarray}
where
\begin{eqnarray}
\Omega_{\tilde{u}}=if_{B_d}\sum_i\frac{\lambda'_{ji3}\lambda_{ki1}'^{*}}{8m^2_{\tilde{u}_{iL}}},~~~
\Omega_{\tilde{\nu}}=-if_{B_d}\mu_{B_d}\sum_i\frac{\lambda_{ijk}\lambda_{i31}'^{*}}{4m^2_{\tilde{\nu}_{iL}}},~~~
\Omega'_{\tilde{\nu}}=if_{B_d}\mu_{B_d}\sum_i\frac{\lambda^*_{ikj}\lambda'_{i13}}{4m^2_{\tilde{\nu}_{iL}}},
\end{eqnarray}
and 
$\mu_{B_d}\equiv\frac{m_{B_d}^2}{\overline{m}_b+\overline{m}_d}$.

Generally, the RPV couplings can be complex and their phase may
induce new contributions, so we write them as
\begin{eqnarray}
\lambda_{ijk}\lambda^*_{lmn} = |\lambda_{ijk}\lambda^*_{lmn}|~e^{i
\phi_{\spur{R_p}}},~~~~~\lambda^*_{ijk}\lambda_{lmn} =
|\lambda_{ijk}\lambda_{lmn}^*|~e^{-i \phi_{\spur{R_p}}},
\end{eqnarray}
and $\phi_{\spur{R_p}} \in [ -\pi, \pi]$ is the RPV weak phase.

%%%%
\subsection{The branching ratios with RPV contributions}

Using the formulae presented above,  we get  the following results
for the double differential decay branching ratios for the decays
$B^+_u\to M^+ \ell^+\ell^-~(M^+=\pi^+~\mbox{or}~\rho^+)$,
\begin{eqnarray}
\frac{d^2\mathcal{B}^{M^+}}{d\hat{s}d\hat{u}}
=\frac{d^2\mathcal{B}^{M^+}_{SM}}{d\hat{s}d\hat{u}}
+\frac{d^2\mathcal{B}^{M^+}_{\tilde{u}}}{d\hat{s}d\hat{u}}+
\frac{d^2\mathcal{B}^{M^+}_{\tilde{\nu}}}{d\hat{s}d\hat{u}}+
\frac{d^2\mathcal{B}'^{M^+}_{\tilde{\nu}}}{d\hat{s}d\hat{u}}.
\end{eqnarray}
Since it is always assumed in the literature  for numerical display that
only one sfermion contributes at one time,  we have neglected
the interferences between different RPV coupling products, but kept
their interferences with the SM amplitude, as shown in the following
equations.

 For the $B^+_u\to \pi^+\ell^+\ell^-$ decay,
\begin{eqnarray}
\frac{d^2\mathcal{B}^{\pi^+}_{\tilde{u}}}{d\hat{s}d\hat{u}}
&=&\tau_B\frac{m^4_B}{2^7\pi^3}
\Bigg\{
Re(WA'\Upsilon^*_{\tilde{u}})(\lambda-\hat{u}^2)\nonumber\\
&&+Re(WC'\Upsilon^*_{\tilde{u}}
)\Big[-(\lambda-\hat{u}^2)-4\hat{m}^2_{\ell}
(2+2\hat{m}^2_{\pi^+}-\hat{s})\Big] \nonumber\\
&&+Re(WD'\Upsilon^*_{\tilde{u}}
)\Big[-4\hat{m}^2_{\ell}(1-\hat{m}^2_{\pi^+})\Big]\nonumber\\
&&+|\Upsilon_{\tilde{u}}|^2m_B\Big[\lambda-\hat{u}^2
+2\hat{m}^2_{\ell}(2+2\hat{m}^2_{\pi^+}-\hat{s})\Big] \Bigg\},
\label{RPVBKmu}
\\
\frac{d^2\mathcal{B}^{\pi^+}_{\tilde{\nu}}}{d\hat{s}d\hat{u}}
&=&\tau_B\frac{m^3_B}{2^7\pi^3}
\Bigg\{
Re(WA'\Upsilon^*_{\tilde{\nu}})(2\hat{m}_{\ell}\hat{u})
+Re(WC'\Upsilon^*_{\tilde{\nu}}
)(1-\hat{m}_{\pi^+}^2)(-2\hat{m}_{\ell})
\nonumber\\
&&+Re(WD'\Upsilon^*_{\tilde{\nu}})(-2\hat{m}_{\ell}\hat{s})
+|\Upsilon_{\tilde{\nu}}|^2(\hat{s}-2\hat{m}^2_{\ell})
\Bigg\},
\label{rpvsnu1}
\\
\frac{d^2\mathcal{B}'^{\pi^+}_{\tilde{\nu}}}{d\hat{s}d\hat{u}}
&=&\tau_B\frac{m^3_B}{2^7\pi^3}
\Bigg\{
Re(WA'\Upsilon'^*_{\tilde{\nu}})(2\hat{m}_{\ell}\hat{u})
+Re(WC'\Upsilon'^*_{\tilde{\nu}}
)(1-\hat{m}_{\pi^+}^2)(2\hat{m}_{\ell})\nonumber\\
&&+Re(WD'\Upsilon'^*_{\tilde{\nu}} )(2\hat{m}_{\ell}\hat{s})
+|\Upsilon'_{\tilde{\nu}}|^2(\hat{s}-2\hat{m}^2_{\ell}) \Bigg\},
\label{rpvsnu2}
\end{eqnarray}
and  $W=-\frac{G_F\alpha_{e}}{2\sqrt{2}~\pi}V^{*}_{tb}V_{td}m_B$.

 For the $B^+_u\to \rho^+\ell^+\ell^-$ decay,
%%%%
\begin{eqnarray}
\frac{d^2\mathcal{B}^{\rho^+}_{\tilde{u}}}{d\hat{s}d\hat{u}}&=&
\tau_B\frac{m_B^3}{2^9\pi^3}
\Bigg\{
Re(WAI^*)
\Big[\hat{s}(\lambda+\hat{u}^2)+4\hat{m}^2_{\ell}\lambda\Big]\nonumber\\
&&-Re(WEI^*)\Big[\hat{s}(\lambda+\hat{u}^2)-4\hat{m}^2_{\ell}\lambda\Big]+|I|^2
\Big[\hat{s}(\lambda+\hat{u}^2)\Big]\nonumber\\
&&+4\hat{s}\hat{u}\Big[Re(WAJ^*)+Re(WBI^*)-Re(WEJ^*)-Re(WFI^*)+2Re
(IJ^*)\Big]\nonumber\\
&&+\frac{1}{\hat{m}^2_{\rho^+}}\Bigg[Re(WBJ^*)\Big(\lambda-\hat{u}^2
+8\hat{m}^2_{\rho^+}(\hat{s}+2m_{\ell}^2)\Big)\nonumber\\
&&-Re(WFJ^*)\Big(\lambda-\hat{u}^2+8\hat{m}^2_{\rho^+}(\hat{s}-4m_{\ell}^2)\Big)\nonumber\\
&&+|J|^2\Big(\lambda-\hat{u}^2+8\hat{m}^2_{\rho^+}(\hat{s}-\hat{m}_{\ell}^2)\Big)\nonumber\\
&&-Re(WBK^*)(\lambda-\hat{u}^2)(1-\hat{m}^2_{\rho^+}-\hat{s})\nonumber\\
&&+Re(WFK^*)\Big((\lambda-\hat{u}^2)(1-\hat{m}^2_{\rho^+}-\hat{s})+4\hat{m}^2_{\ell}\lambda\Big)
\nonumber\\
&&-2Re(JK^*)\Big((\lambda-\hat{u}^2)(1-\hat{m}^2_{\rho^+}-\hat{s})+2\hat{m}^2_{\ell}\lambda\Big)\Bigg]
\nonumber\\
&&+\frac{\lambda}{\hat{m}^2_{\rho^+}}\Bigg[Re(WCK^*)(\lambda-\hat{u}^2)-Re(WGK^*)
\Big(\lambda-\hat{u}^2+4\hat{m}^2_{\ell}(2+2\hat{m}^2_{\rho^+}-\hat{s})\Big)\nonumber\\
&&+|K|^2\Big(\lambda-\hat{u}^2+2\hat{m}^2_{\ell}(2+2\hat{m}^2_{\rho^+}-\hat{s})\Big)\Bigg]\nonumber\\
&&+\frac{4\hat{m}^2_{\ell}}{\hat{m}^2_{\rho^+}}\lambda\Bigg[-Re(WHL^*)\hat{s}+|L|^2\hat{s}/2
+Re(WFL^*)-Re(JL^*)\nonumber\\
&&-Re(WGL^*)(1-\hat{m}^2_{\rho^+})+Re(KL^*)(1-\hat{m}^2_{\rho^+})
 \Bigg]
    \Bigg\}, \\
\frac{d^2\mathcal{B}^{\rho^+}_{\tilde{\nu}}}{d\hat{s}d\hat{u}}&=&\tau_B\frac{m^3_B}{2^7\pi^3}
\Bigg\{-\frac{\hat{m}^2_{\ell}}{\hat{m}^2_{\rho^+}}
\Bigg[Im(WB\Phi_{\tilde{\nu}}^*)\cdot
\Big(\lambda^{-\frac{1}{2}}\hat{u}(1-\hat{m}^2_{\rho^+}-\hat{s})\Big)\nonumber\\
&&+Im(WC\Phi_{\tilde{\nu}}^*)\lambda^{\frac{1}{2}}\hat{u}
+Im(WF\Phi_{\tilde{\nu}}^*)\lambda^{\frac{1}{2}}\nonumber\\
&&-Im(WG\Phi_{\tilde{\nu}}^*)\lambda^{\frac{1}{2}}(1-\hat{m}^2_{\rho^+})\Bigg]
+|\Phi_{\tilde{\nu}}|^2(\hat{s}-2\hat{m}^2_{\ell})\Bigg\},\\
\frac{d^2\mathcal{B}'^{\rho^+}_{\tilde{\nu}}}{d\hat{s}d\hat{u}}&=&\tau_B\frac{m^3_B}{2^7\pi^3}
\Bigg\{-\frac{\hat{m}^2_{\ell}}{\hat{m}^2_{\rho^+}}
\Bigg[Im(WB\Phi'^*_{\tilde{\nu}})\cdot
\Big(\lambda^{-\frac{1}{2}}\hat{u}(1-\hat{m}^2_{\rho^+}-\hat{s})\Big)\nonumber\\
&&+Im(WC\Phi'^*_{\tilde{\nu}})
\lambda^{\frac{1}{2}}\hat{u}-Im(WF\Phi'^*_{\tilde{\nu}})\lambda^{\frac{1}{2}}\nonumber\\
&&+Im(WG\Phi'^*_{\tilde{\nu}})\lambda^{\frac{1}{2}}(1-\hat{m}^2_{\rho^+})\Bigg]
+|\Phi'_{\tilde{\nu}}|^2(\hat{s}-2\hat{m}^2_{\ell})\Bigg\}.
\end{eqnarray}

The normalized forward-backward asymmetries is defined by
\begin{eqnarray}
\mathcal{A}_{FB}(B\to M\ell^+\ell^-)=\int
d\hat{s}~\frac{\int^{+1}_{-1}\frac{d^2\mathcal{B}(B\to
M\ell^+\ell^-)}{d\hat{s}dcos\theta}sign(cos\theta)dcos\theta}
{\int^{+1}_{-1}\frac{d^2\mathcal{B}(B\to
M\ell^+\ell^-)}{d\hat{s}dcos\theta}dcos\theta}.
\end{eqnarray}
 Since there is no term containing $\hat{u}$ with an odd power as shown by Eq.(\ref{Bpi}), the
$\mathcal{A}_{FB}$ is zero for  $B^+_u\to \pi^+\ell^+\ell^-$
decays in the SM. The RPV effect via  squark exchange on
$\mathcal{A}_{FB}(B^+_u\to\pi^+\ell^+\ell^-)$ also vanishes for the
same reason as shown by Eq.(\ref{RPVBKmu}), while  the
sneutrino exchange contributions to $\mathcal{A}_{FB}$, as shown by
Eqs.(\ref{rpvsnu1}, \ref{rpvsnu2}), are proportional to $m_{\ell}$,
which are tiny for $\ell=e, \mu$.

The total decay branching ratios of the pure leptonic $B^0_d$ decays
are given by
\begin{eqnarray}
\mathcal{B}(B_d\to \ell^+\ell^-)&=&\mathcal{B}^{SM}(B_d\to
\ell^+\ell^-)\Bigg\{1+\frac{1}{|K_{SM}|^2}\bigg[2Re(K_{SM}\Omega^*_{\tilde{u}})
+|\Omega_{\tilde{u}}|^2
\bigg]\nonumber\\
&&+\frac{1}{|K_{SM}|^2}\bigg[Re(K_{SM}\Omega^*_{\tilde{\nu}})\frac{1}{m_{\ell}}+
|\Omega_{\tilde{\nu}}|^2
\bigg(\frac{1}{2m^2_{\ell}}-\frac{1}{m^2_{B_d}}\bigg)\bigg]\nonumber\\
&&+\frac{1}{|K_{SM}|^2}\bigg[-Re(K_{SM}\Omega^{'*}_{\tilde{\nu}})\frac{1}{m_{\ell}}+
|\Omega^{'}_{\tilde{\nu}}|^2
\bigg(\frac{1}{2m^2_{\ell}}-\frac{1}{m^2_{B_d}}\bigg)\bigg]\Bigg\},
\label{btoll}
\end{eqnarray}
with
\begin{eqnarray}
K_{SM}=-\frac{G_F}{\sqrt{2}}\frac{\alpha_{e}}{2\pi\mbox{sin}2\theta_W}V_{td}V^{*}_{tb}Y(x_t)(if_{B_d}).
\end{eqnarray}
From Eq.(\ref{btoll}), we can see that $\mathcal{B}(B^0_d\to
\ell^+\ell^-)$ could be enhanced very much by the s-channel RPV
sneutrino exchange, but not so much by the t-channel squark exchange.

\section{Numerical results and analyses}

With the formulae  presented in previous section, we are ready to
perform our numerical analysis.  Firstly, we will specify the input
parameters.  Then we will use the latest up-limits of
$\mathcal{B}(B^{+}_{u}\to \pi^{+}\ell^+\ell^-)$ and
 $\mathcal{B}(B^{0}_{d}\to\ell^+\ell^-)$ to get the constraints on the
 relevant RPV couplings. Finally, we will show the distributions of
 the branching ratios and the forward-backward
asymmetries  of  these decays in the constrained  RPV parameter
space.

In the Wolfenstein parametrization of the CKM matrix elements, one has
 \begin{eqnarray}
\lambda_{u}^*=\frac{V^{*}_{ub}V_{ud}}{V^{*}_{tb}V_{td}}
 =\frac{\bar{\rho}+i\bar{\eta}}{1-\bar{\rho}-i \bar{\eta}}+O(\lambda^5).
\end{eqnarray}

The parameters $ \lambda, \bar{\rho}$ and $\bar{\eta}$ have been
updated by CKMfitter   Group \cite{CKMfactor}, which read
\begin{eqnarray}
\lambda=0.22568^{+0.00084}_{-0.00079},~~~~\bar{\rho}=0.213^{+0.014}_{-0.024},
~~~~\bar{\eta}=0.348^{+0.012}_{-0.021}.
\end{eqnarray}
In our numerical analysis, we have used
$|V_{td}V_{tb}^{*}|=|V_{td}|=(8.27^{+0.21}_{-0.49})\times 10^{-3}$
\cite{CKMfactor}.

 For the semileptonic decays
$B^+_u\to\pi^+\ell^+\ell^-$ and $B^+_u\to\rho^+\ell^+\ell^-$,   we
will use the form factors  of light-cone QCD sum rules (LCSRs)
results \cite{BallZwicky}, which are renewed  recently with
radiative corrections to the leading twist wave functions and SU(3)
breaking effects. For the $q^2$ dependence of the form factors, they
are parameterized in terms of simple formulae with two or three
parameters. The form factors $V, ~ A_{0}, ~T_{1}, f_{+}$ and $f_T$
are parameterized by  \cite{BallZwicky}
\begin{eqnarray}
F(\hat{s})&=&\frac{r_1}{1-\hat{s}/\hat{m}^2_R}+\frac{r_2}{1-\hat{s}/\hat{m}^2_{fit}}.\label{r12mRfit}
\end{eqnarray}
For the form factors $A_2$ and $\tilde{T}_3$, it is more appropriate
to  expand them  to the second order around the pole
\cite{BallZwicky}
\begin{eqnarray}
F(\hat{s})&=&\frac{r_1}{1-\hat{s}/\hat{m}^2_{fit}}+\frac{r_2}{(1-\hat{s}/\hat{m}^2_{fit})^2}.\label{r12mfit}
\end{eqnarray}
The fit formula for $A_1$, $T_2$ and $f_0$ is
\begin{eqnarray}
F(\hat{s})&=&\frac{r_2}{1-\hat{s}/\hat{m}^2_{fit}}.\label{r2mfit}
\end{eqnarray}
The form factor $T_3$ can be obtained with
$T_3(\hat{s})=\frac{1-\hat{m}_{\rho}}{\hat{s}}[\widetilde{T}_3(\hat{s})-T_2(\hat{s})].$
The values of all the corresponding parameters for these form
factors are listed in Table I, the uncertainties induced by F(0)
\cite{BallZwicky} are also considered in the following numerical
data analyses.
\begin{table}[h]
\centerline{\parbox{15cm}{\small Table I: Fit for form factors
involving the $B\to \pi(\rho)$ transitions valid for general $q^2$
 \cite{BallZwicky}.}} \vspace{0.3cm}
\begin{center}
\begin{tabular}{cccccccc}\hline\hline
$F(\hat{s})$&$~F(0)~$&$~\Delta_{tot}~$&$~r_1~$&$~m_R^2~$&$~r_2~$&$~m^2_{fit}~$&~fit
Eq.\\\hline $f_+^{B\to
\pi}$&$0.258$&$0.031$&$0.744$&$5.32^2$&$-0.486$&$40.73$&(\ref{r12mRfit})\\\hline
$f_T^{B\to
\pi}$&$0.253$&$0.028$&$1.387$&$5.32^2$&$-1.134$&$32.22$&(\ref{r12mRfit})\\\hline
$f_0^{B\to
\pi}$&$0.258$&$0.031$&&&$0.258$&$33.81$&(\ref{r2mfit})\\\hline
$V^{B\to
\rho}$&$0.323$&$0.030$&$1.045$&$5.32^2$&$-0.721$&$38.34$&(\ref{r12mRfit})\\\hline
$A_0^{B\to
\rho}$&$0.303$&$0.029$&$1.527$&$5.28^2$&$-1.220$&$33.36$&(\ref{r12mRfit})\\\hline
$A_1^{B\to
\rho}$&$0.242$&$0.023$&$$&$$&$0.240$&$37.51$&(\ref{r2mfit})\\\hline
$A_2^{B\to
\rho}$&$0.221$&$0.023$&$0.009$&$$&$0.212$&$40.82$&(\ref{r12mfit})\\\hline
$T_1^{B\to
\rho}$&$0.267$&$0.023$&$0.897$&$5.32^2$&$-0.629$&$38.04$&(\ref{r12mRfit})\\\hline
$T_2^{B\to
\rho}$&$0.267$&$0.023$&$$&$$&$0.267$&$38.59$&(\ref{r2mfit})\\\hline
$\widetilde{T}_3^{B\to
\rho}$&$0.267$&$0.024$&$0.022$&$$&$0.246$&$40.88$&(\ref{r12mfit})\\\hline
\end{tabular}
\end{center}
\end{table}

The values of other input parameters used in the numerical
calculation and the experimental upper-limits  are given, respectively,  in Tables II and
III. In the numerical calculations through the paper, we will  take into account of the $1\sigma$
uncertainties of the parameters with relative large error bars.

With these inputs, we get the SM prediction 
 $\mathcal{B}(B^+_u \to \pi^+ \ell^+\ell^-) =(2.03\pm 0.23)\times10^{-8}$, which is smaller than the result
($3.27\times10^{-8}$) in Ref. \cite{Aliev2}. The difference is  mainly due to the updated values for formfactors 
and  CKM parameters.

 As in the literature, we assume that only one sfermion
contributes at one time with a mass of 100 GeV.   For other values
of the sfermion masses, the bounds on the couplings in this paper
can be easily obtained by scaling them by factor
$\tilde{f}^2\equiv(\frac{m_{\tilde{f}}}{100GeV})^2$.
%
%%%%%%%%%%%%table II %%%%%%%%%%%%%%%%%%%%%%%
\begin{table}[h]
\centerline{\parbox{15cm}{\small Table II: Default values of the
input parameters and the $\pm1 \sigma$ error bars of  the sensitive
parameters used in our numerical calculations in addition to the
ones discussed in text.}} \vspace{0.3cm}
\begin{center}
\begin{tabular}{lc}\hline\hline
$m_{B_d}=5.279~GeV,~~m_{B_u}=5.279~GeV,~~m_W=80.425~GeV,$& \\
$m_{\pi^+}=0.140~GeV,~~m_{\pi^0}=0.135~GeV,~~m_\rho=0.776~GeV,$& \\
$\overline{m}_b(\overline{m}_b)=(4.20\pm0.07)~GeV,$& \\
$\overline{m}_u(2GeV)=0.0015\sim
0.003~GeV,~\overline{m}_d(2GeV)=0.003\sim
0.007~GeV,$& \\
$m_e=0.511\times10^{-3}~GeV,~~m_\mu=0.106~GeV,$~~
$m_{t,pole}=174.2\pm3.3~GeV. $&  \cite{PDG}\\ \hline
$\tau_{B_{d}}=1.530~ps,~~\tau_{B_{u}}=1.638~ps.$&  \cite{PDG}\\
\hline $\mbox{sin}^2\theta_W=0.22306,~~\alpha_e=1/137.$&
 \cite{PDG}\\\hline $f_{B_d}=0.216^{+0.009}_{-0.019}~ GeV.$&
 \cite{fBd}\\\hline\hline
\end{tabular}
\end{center}
\end{table}
%%%%%%%%%%%Table III %%%%%%%%%%
\begin{table}[h]
\centerline{\parbox{14cm}{\small Table III: The SM predictions and
experimental data of branching ratios for the upper limits for
$B^0_d\to\ell^+\ell^-$ and $B^+_u\to\pi^+\ell^+\ell^-$
 \cite{Aubert,cdf,purebf}.}} \vspace{0.4cm}
\begin{center}
\begin{tabular}
{l|c|c}\hline\hline & SM prediction value & Experimental
data\\\hline
 $\mathcal{B}(B^+_u \to \pi^+\mu^+\mu^-)$&$(2.03\pm 0.23) \times10^{-8}$ &
 $<2.8\times10^{-7}$~(90\%$~\emph{C.} \emph{L.}$)  \cite{Aubert}\\
$\mathcal{B}(B^+_u \to \pi^+ e^+e^-)$&$(2.03\pm
0.23)\times10^{-8}$&$<1.8\times10^{-7}$~(90\%$~\emph{C.} \emph{L.}$)
 \cite{Aubert}\\
 $\mathcal{B}(B^+_u \to \rho^+\mu^+\mu^-)$&$(4.33\pm 1.14)\times10^{-8}$&$$\\
 $\mathcal{B}(B^+_u \to \rho^+ e^+e^-)$& $(5.26\pm 1.37)\times10^{-8}$&$$\\
 $\mathcal{B}(B_d\to \mu^+\mu^-)$& $(1.33\pm 0.12)\times10^{-10}$&$<1.5\times10^{-8}
 $~(90\%$~\emph{C.} \emph{L.}$) \cite{cdf} \\
 $\mathcal{B}(B_d \to e^+e^-)$&$(3.11\pm 0.27)\times10^{-15}$&$<6.1\times10^{-8}
 $~(90\%$~\emph{C.} \emph{L.}$)  \cite{purebf}\\ \hline\hline
\end{tabular}
\end{center}
\end{table}
%%%%%%%%%%%%%%%
We present the branching ratios of  the SM  expectation  in Table
III,   where  the relevant experimental upper limits \cite{Aubert,
cdf, purebf} are listed  for comparison. We can see that the
experimental upper limits of  $\mathcal{B}(B^+_u \to
\pi^+\ell^+\ell^-) $ are just one order above the SM expectations,
which will constrain RPV SUSY parameter space significantly.
Although the present limit of $\mathcal{B}(B^{0}_d\to \mu^+\mu^-)$
are three orders of magnitudes higher than the SM predictions,  the
upper-limit will also yield strong bounds since the decay
$B^{0}_d\to \mu^+\mu^-  $ is very sensitive to new pseudo-scalar
interactions beyond the SM.
%%%%%Fig.\ref{bounds}.%%%%%%
\begin{figure}[h]
\begin{center}
\includegraphics[scale=1.4]{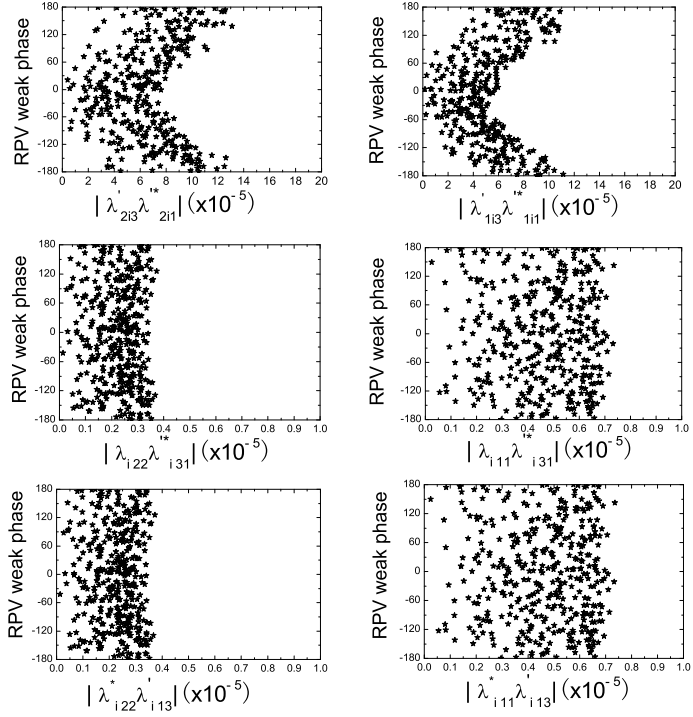}
\end{center}
%\vspace{-0.6cm}
 \caption{ The allowed parameter spaces for the relevant
 RPV couplings constrained by the measurements  listed in Table IV,
 and the RPV weak phase is given in degree.}
 \label{bounds}
\end{figure}
%%%%%%%%%%%

Now we turn to the RPV effects. The decays to
$B^+_u\to\rho^+\ell^+\ell^-$, $B^+_u\to\pi^+\ell^+\ell^-$ and
$B^{0}_d\to \ell^+\ell^-$ involve the same set of the
 six RPV coupling products.    Combining  the upper-limits of
 $\mathcal{B}(B^+_u \to \pi^+\ell^+\ell^-) $ \cite{Aubert} and
 $\mathcal{B}(B^{0}_d\to \ell^+\ell^-)$  \cite{cdf, PDG}  listed in Table III,
we get constraints on the six RPV coupling products. Subject to the
constraints,  the random variations of the parameters as discussed
previously lead  to the scatter plots displayed in Fig.\ref{bounds}.
From the figure, we can see that  the constraints RPV on
$\lambda^{\prime}_{2i3} \lambda^{\prime\ast}_{2i1}$ and
  $\lambda^{\prime}_{1i3} \lambda^{\prime\ast}_{1i1}$ are sensitive to their RPV weak phases, which  implies  significant  interferences
  between the squark contributions and the SM ones. As depicted  by Fig.1 and 2,
  the squark contribution amplitudes are of $\bar{b}(V+A)d \bar{\ell}(V-A) \ell$
  current structure after Fierz transformations, which are similar  to the SM amplitudes.
  For the constraints on  $\lambda^{\prime} \lambda^{\ast}$ of sneutrinos  are insensitive  to
 their phase $\phi_{\spur{R_p}}$.   We summarize  all the constraints in Table IV, where the  recent updated
  bounds \cite{dreiner} are listed for comparison.  It's shown  that our bounds
  are more restricted than the previous ones.
   We note that the strong constraints on $\lambda'_{1i3}\lambda'^{*}_{1i1}$ and
$\lambda'_{2i3}\lambda'^{*}_{2i1}$ are mainly due to the
upper-limits of  $\mathcal{B}(B^{+}\to \pi^{+}\ell^{+}\ell^{-})$,
which are just one order higher that the SM expectations,
and   the constraints on other four RPV coupling products
from the sneutrino exchange are mainly due to the upper-limits of
$\mathcal{B}(B^{0}_d\to\ell^{+}\ell^{-})$.
%%%%%%%%%%%%%%%TABLE%%%%%%%%%%%%
\begin{table}[h]
\centerline{\parbox{13.8cm}{Table IV: \small {Bounds for  the
relevant RPV couplings products by $B^+_u \to \pi^+\ell^+\ell^-$ and
$B^{0}_d \to \ell^+\ell^-$ decays for 100 $GeV$ sfermions, and
previous bounds are listed for comparison. }}} \vspace{0.5cm}
\begin{center}
\begin{tabular}{c|l|l}\hline\hline
Couplings&~~~~~~~~~Bounds [Processes]& Previous bounds [Processes]\cite{dreiner}\\
\hline $|\lambda'_{1i3}\lambda'^*_{1i1}|$&$\leq1.1\times
10^{-4}~_{[B^+_u\to \pi^+ e^+e^-]}^{[B_d \to e^+e^-]}$ &$\leq2.6\times 10^{-2}~[B_d \to e^+e^-]$  \\
$|\lambda_{i11}\lambda'^*_{i31}| $&$\leq7.4\times 10^{-6}~_{[
B^+_u\to
\pi^+ e^+e^-]}^{[B_d \to e^+e^-]}$&$\leq4.1\times 10^{-5}~[B_d \to e^+e^-]$  \\
$|\lambda^*_{i11}\lambda'_{i13}|$&$\leq7.4\times 10^{-6}~_{[B^+_u\to
\pi^+ e^+e^-]}^{[B_d \to e^+e^-]}$&$\leq4.1\times 10^{-5}~[B_d \to e^+e^-]$  \\
$|\lambda'_{2i3}\lambda'^*_{2i1}| $&$\leq1.3\times
10^{-4}~_{[B^+_u\to \pi^+\mu^+\mu^-]}^{[B_d \to
\mu^+\mu^-]}$&$\leq5.4\times
10^{-4}~[B_d \to\mu^+\mu^-]$  \\
$|\lambda_{i22}\lambda'^*_{i31}| $&$\leq3.7\times
10^{-6}~_{[B^+_u\to \pi^+\mu^+\mu^-]}^{[B_d \to
\mu^+\mu^-]}$&$\leq6.2\times
10^{-6}~[B_d\to\mu^+\mu^-]$  \\
$|\lambda^*_{i22}\lambda'_{i13}| $&$\leq3.7\times
10^{-6}~_{[B^+_u\to \pi^+\mu^+\mu^-]}^{[B_d \to
\mu^+\mu^-]}$&$\leq6.2\times 10^{-6}~[B_d\to \mu^+\mu^-]$
 \\\hline\hline
\end{tabular}
\end{center}
\end{table}
%%%%%%%%%%%%%%%%%%%%%%%%%%%%%%%%%%%%

Using the constrained  parameter spaces shown in Table IV and
Fig.\ref{bounds}, one can predict the RPV effects on the other
quantities which have not been well measured yet in these processes.
We perform a scan over the input parameters and the new constrained
RPV coupling spaces to get the allowed ranges for
$\mathcal{B}(B^+_u\to\rho^+\ell^+\ell^-)$,
$\mathcal{B}(B^+_u\to\pi^+\ell^+\ell^-)$,
$\mathcal{B}(B^{0}_d\to\ell^+\ell^-)$,
$\mathcal{A}_{FB}(B^+_u\to\pi^+\ell^+\ell^-)$ and
$\mathcal{A}_{FB}(B^+_u\to \rho^+\ell^+\ell^-)$ with
 the different RPV couplings products.  The numerical results are
summarized  in Table V.
%%%
\begin{table}[htbp]
\centerline{\parbox{16.6cm}{\small Table V: The results for $\mathcal{B}(B^0_d\to \ell^+\ell^-)$,
$\mathcal{B}(B^+_u \to\rho^+\ell^+\ell^-, \pi^+\ell^+\ell^-)$ and
$\mathcal{A}_{FB}(B^+_u\to \pi^+(\rho^+)\ell^+\ell^-)$ in the SM and
the RPV SUSY. The RPV SUSY predictions are obtained by the allowed
regions of the different RPV couplings products. In the RPV
coulpings, $g=1$ and $2$ for $\ell=e$ and $\mu$, respectively.}}
\vspace{0.3cm}
\begin{center}\footnotesize{
\begin{tabular}{l|c|c|c|c}\hline\hline
&SM value
&$\lambda'_{gi3}\lambda'^*_{gi1}$&$\lambda_{igg}\lambda'^*_{i31}$&$\lambda^*_{igg}\lambda'_{i13}$\\\hline
$\mathcal{B}(B^0_d\to
e^+e^-)$&$[2.2,3.7]\times10^{-15}$&$[0.002,6.7]\times 10^{-14}$
&$[0.008,6.1]\times10^{-8}$&$[0.008,6.1]\times10^{-8}$\\\hline
$\mathcal{B}(B^+_u\to\pi^+
e^+e^-)$&$[1.4,2.6]\times10^{-8}$&$[0.57,18.0]\times10^{-8}$&$[1.6,15.5]\times10^{-8}$
&$[1.7,13.8]\times10^{-8}$\\\hline $\mathcal{B}(B^+_u\to\rho^+
e^+e^-)$&$[2.8,9.7]\times10^{-8}$&$[3.8,586.8]\times10^{-8}$&$[2.9,10.2]\times10^{-8}$
&$[2.8,10.2]\times10^{-8}$\\\hline $\mathcal{A}_{FB}(B^+_u\to
\pi^+e^+e^-)$&$0$&$0$
&$[-7.2,6.7]\times10^{-5}$&$[-7.2,7.2]\times10^{-5}$\\\hline
$\mathcal{A}_{FB}(B^+_u\to \rho^+
e^+e^-)$&$[0.14,0.23]$&$[-0.13,0.82]$ &$[0.14,0.23]$&$[0.15,0.23]$
\\\hline
$\mathcal{B}(B^0_d\to
\mu^+\mu^-)$&$[0.96,1.56]\times10^{-10}$&$[0.01,38.9]\times10^{-10}$
&$[0.01,1.5]\times10^{-8}$&$[0.01,1.5]\times10^{-8}$\\\hline
$\mathcal{B}(B^+_u\to\pi^+\mu^+\mu^-)$&$[1.4,2.6]\times10^{-8}$&$[0.22,27.92]\times10^{-8}$
&$[1.7,5.3]\times10^{-8}$&$[1.6,5.4]\times10^{-8}$\\\hline
$\mathcal{B}(B^+_u\to\rho^+\mu^+\mu^-)$&$[2.8,8.4]\times10^{-8}$&$[3.5,780.4]\times10^{-8}$
&$[2.8,9.1]\times10^{-8}$&$[2.8,9.1]\times10^{-8}$\\\hline
 $\mathcal{A}_{FB}(B^+_u\to \pi^+\mu^+\mu^-)$&$0$&$0$
&$[-1.1,1.1]\times10^{-2}$&$[-1.1,1.1]\times10^{-2}$\\\hline
$\mathcal{A}_{FB}(B^+_u\to
\rho^+\mu^+\mu^-)$&$[0.14,0.23]$&$[-0.13,0.46]$
&$[0.14,0.23]$&$[0.14,0.24]$
\\\hline \hline
\end{tabular}}
\end{center}
\end{table}
%
%
%%%%%%%% FIGs %%%4%%%%%%%%%%
\begin{figure}[ht]
\begin{center}
\includegraphics[scale=0.7]{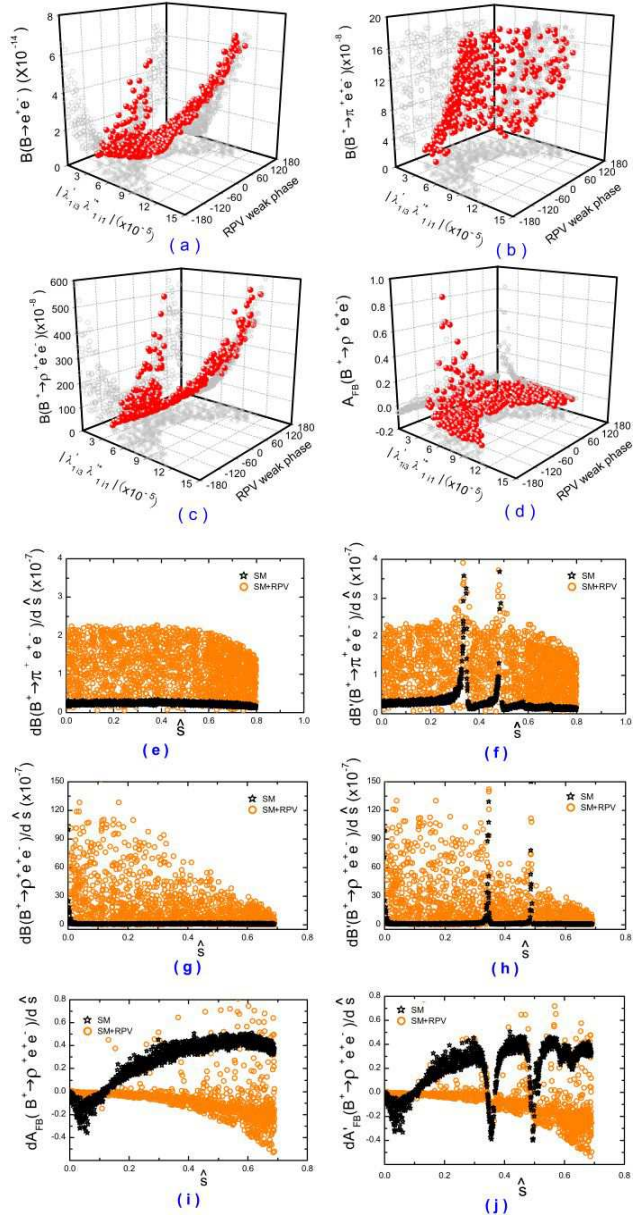}
\end{center}
\caption{ The effects of RPV coupling $\lambda'_{1i3}
 \lambda'^*_{1i1}$ due to the squark exchange
 in $B^+_u\to \pi^+e^+e^-$, $B^+_u\to \rho^+e^+e^-$ and $B^0_d\to e^+e^-$ decays.
 The primed observables are given with $\psi(nS)$
 VMD contributions.} 
 \label{figelplps}
\end{figure}
%Fig4%%
%
\begin{figure}[ht]
\begin{center}
\includegraphics[scale=0.7]{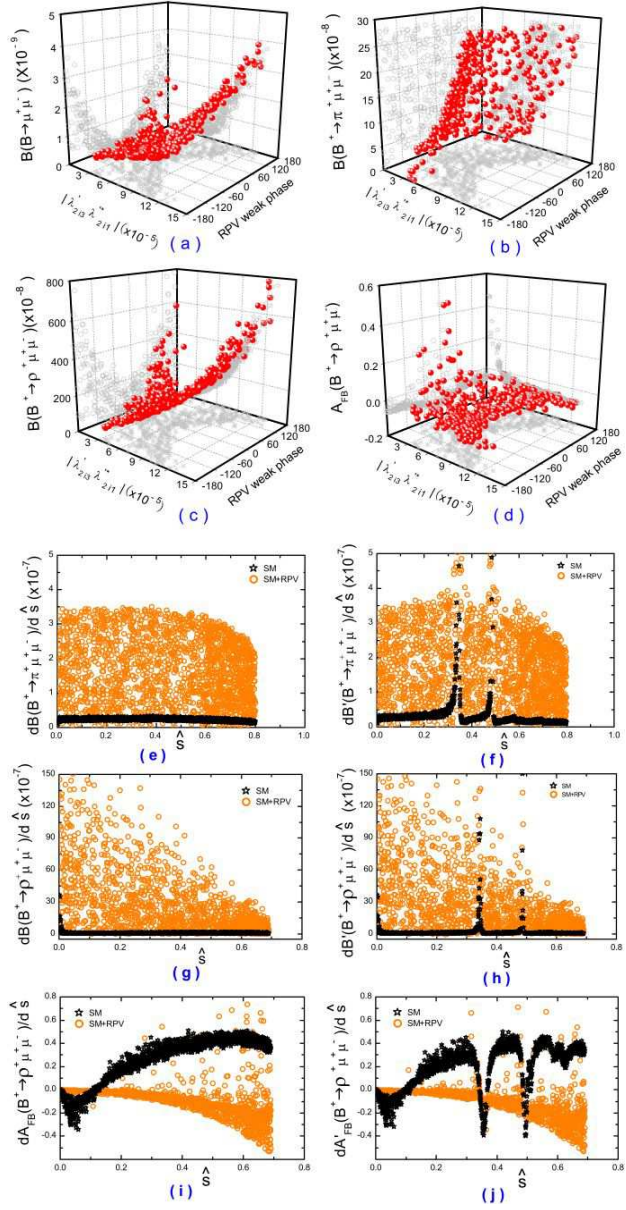}
\end{center}
\caption{ The effects of RPV coupling $\lambda'_{2i3}
 \lambda'^*_{2i1}$ due to the squark exchange in $B^+_u\to
 \pi^+\mu^+\mu^-$, $B^+_u\to\rho^+\mu^+\mu^-$
 and $B^0_d\to \mu^+\mu^-$ decays.}
 \label{figulplps}
\end{figure}
%
% Fig5 %%
%%%%%%%% FIGs %%%%%%%%%%%%%

For  Table V,  we have the following remarks.
%%%%
%%%%
\begin{itemize}
 \item   As shown by  Fig.\ref{purefig}(b), the contributions of
$\lambda'_{1i3}\lambda'^{*}_{1i1}$ and
$\lambda'_{2i3}\lambda'^{*}_{2i1}$ to $\mathcal{B}(B^0_d\to e^+e^-)$
and $\mathcal{B}(B^0_d\to\mu^+\mu^-)$, respectively, arise from
t-channel squark exchange. After Fierz transformation, the effective
Hamiltonian due to the t-channel squrk exchange is proportional to
$\bar{b}\gamma^{\mu}P_{R}d \bar{\ell}\gamma_{\mu}P_{L}\ell$, which
contributions to $\mathcal{B}(B^0_d\to \ell^+\ell^-)$ are suppressed
by $m^2_{\ell}/m^2_B$ due to helicity suppression. Therefore,
$\mathcal{B}(B^0_d\to e^+e^-, \mu^+\mu^-)$ will not be enhanced so
much by the t-channel squark exchanging RPV contributions.  In other
words, the bounds on these two RPV coupling products are due to the
upper-limits of $\mathcal{B}( B^+_u\to \pi^+\ell^+\ell^- ) $. With
the bounds, the RPV squark exchanges could enhance
 $\mathcal{B}( B^{0}_{d} \to e^{+}e^{-})$ and
  $\mathcal{B}( B^{0}_{d} \to \mu^{+} \mu^{-})$,  to the utmost, to $6.7\times 10^{-14}$
   and $3.9\times 10^{-9} $,  respectively. However, the effective Hamiltonian of s-channel sneutrino exchange
would be $\bar{b}(1\pm\gamma_5)d \bar{\ell}(1\mp\gamma_5)\ell$,
whose contributions are not suppressed by $m^2_{\ell}/m^2_B$.
Therefore, both $\mathcal{B}(B^0_d\to e^+e^-)$ and
$\mathcal{B}(B^0_d\to \mu^+\mu^-)$ could be enhanced to order of 
$10^{-8}$ to saturate their experimental upper limits by
$\lambda_{i11}\lambda_{i31}'^{*}$ and $\lambda_{i22}\lambda_{i31}'^{*}$,
respectively.

\item
At present there is no experimental data for  $B^{+}_u\to
\rho^{+}\ell^{+}\ell^{-}$ decays.
 It is interesting to note that  $\mathcal{B}(B^{+}_u\to \rho^{+} \ell^{+}\ell^{-} )$
  could be enhanced to order of $10^{-6}$ by
 the $t$-channel squark exchange within the bounds  due to the upper-limits of
 $\mathcal{B}(B^{+}_u\to \pi^{+} \ell^{+}\ell^{-} )$,  however,
 it is not sensitive to the $s$-channel sneutrino exchanges within the constrained parameter space.

\item  $\mathcal{A}_{FB}(B^+_u\to\pi^+e^+e^-)$ and
$\mathcal{A}_{FB}(B^+_u\to\pi^+\mu^+\mu^-)$ are zero in the SM. The
RPV contributions to the asymmetries due to squark exchanging are
also zero, while the sneutino exchanging RPV contributions are also
too small to be accessible at the LHC.

\item It is found  that the squark exchanging RPV contributions
have significant impacts on $\mathcal{A}_{FB}(B^+_u\to\rho^+e^+e^-)$
and $\mathcal{A}_{FB}(B^+_u\to\rho^+\mu^+\mu^-)$.
 While, the RPV contributions due to the sneutrino
exchange affect  $\mathcal{A}_{FB}(B^+_u\to \rho^+e^+e^-)$ and
$\mathcal{A}_{FB}(B^+_u\to \rho^+\mu^+\mu^-)$ slightly.
\end{itemize}
%%%%%
%%%%%

\begin{figure}[h]
\begin{center}
\includegraphics[scale=0.8]{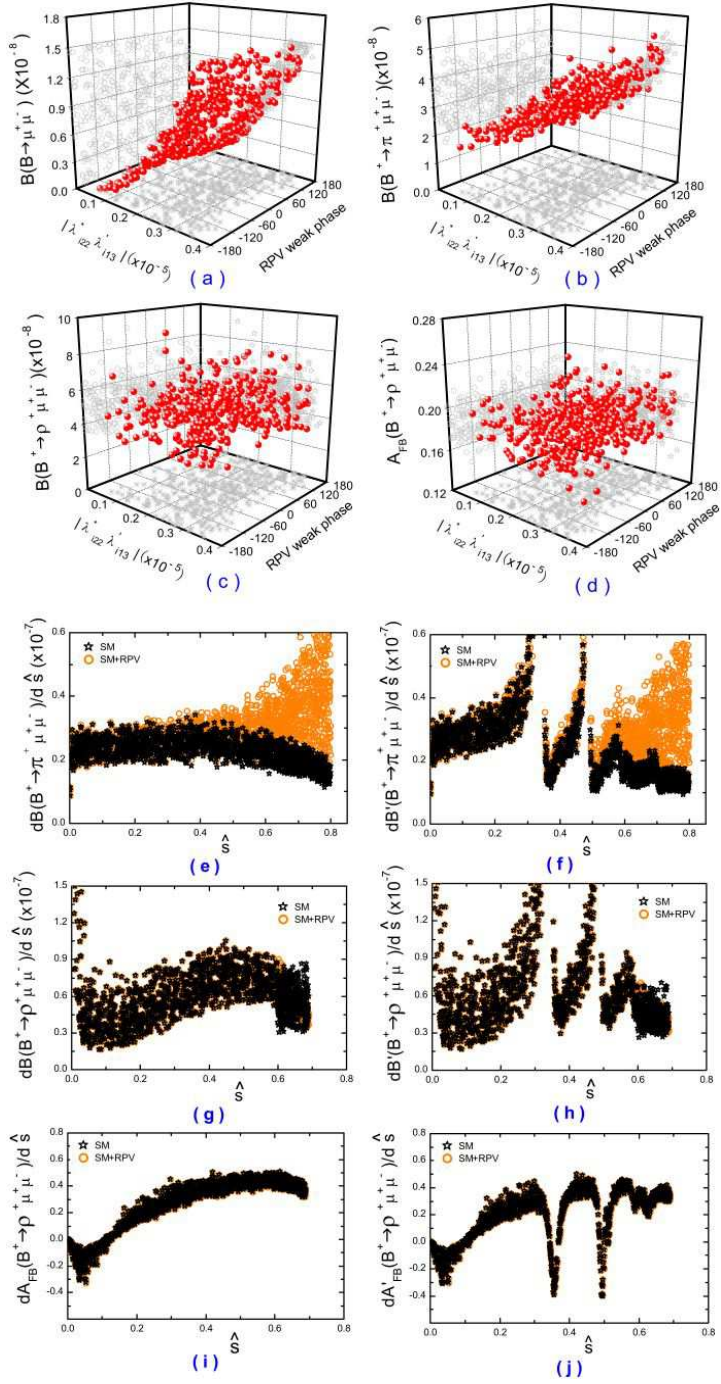}
\end{center}
\caption{The effects of RPV coupling $\lambda^*_{i22}
 \lambda'_{i13}$ due to the sneutrino exchange in $B^+_u\to
 \pi^+\mu^+\mu^-$, $B^+_u\to \rho^+\mu^+\mu^-$
 and $B^0_d\to \mu^+\mu^-$ decays.
 }\label{figulslp}
\end{figure}

For each RPV coupling product,  we present the distributions and
correlations of branching ratios and forward-backward asymmetries
within the constrained parameter space in Fig.\ref{bounds}.
Numerical results are shown in Figs.\ref{figelplps}-\ref{figulslp},
where  correlations between the physical observable $\mathcal{B}$,
$\mathcal{A}_{FB}$ and the parameter spaces of the different RPV
coupling products by the three-dimensional scatter plots. The
dilepton invariant mass distribution and the normalized
forward-backward asymmetry are given with vector meson dominance
(VMD) contribution excluded in terms of $d\mathcal{B}/d\hat{s}$ and
$d\mathcal{A}_{FB}/d\hat{s}$, and included in
$d\mathcal{B}'/d\hat{s}$ and $d\mathcal{A}'_{FB}/d\hat{s}$,
respectively. From Figs.\ref{figelplps}-\ref{figulslp}, one can find
the correlations of these physical observable with the RPV coupling
products.
%%%Fig6%%

Now we turn to discuss plots of Fig.\ref{figelplps} in detail. The
three-dimensional scatter plot shows $\mathcal{A}_{FB}$ and
$\mathcal{B}$ correlated with $|\lambda'_{1i3}\lambda'^*_{1i1}|$ and
its phase $\phi_{\spur{R_p}}$. We also give projections on three
vertical planes, where the
$|\lambda'_{1i3}\lambda'^*_{1i1}|$-$\phi_{\spur{R_p}}$ plane
displays the constrained regions of $\lambda'_{1i3}
\lambda'^*_{1i1}$ as the second plot of Fig.\ref{bounds}. From plots
of Fig.\ref{figelplps}(a)-(c), we can find that
$\mathcal{B}(B^0_d\to e^+e^-)$, $\mathcal{B}(B^+_u\to \pi^+e^+e^-)$
and $\mathcal{B}(B^+_u\to \rho^+e^+e^-)$ are increasing with
$|\lambda'_{1i3}\lambda'^*_{1i1}|$ on the
$\mathcal{B}$-$|\lambda'_{1i3}\lambda'^*_{1i1}|$ planes, and
$\mathcal{B}(B^+_u\to \pi^+e^+e^-)$ can be enhanced to its present
experimental upper limit. Fig.\ref{figelplps}(d) shows that
$\mathcal{A}_{FB}(B^+_u\to \rho^+e^+e^-)$ is decreasing with
$|\lambda'_{1i3}\lambda'^*_{1i1}|$ on the
$\mathcal{A}_{FB}(B^+_u\to\rho^+e^+e^-)$-$|\lambda'_{1i3}
\lambda'^*_{1i1}|$ plane. It is shown that $\mathcal{B}(B^0_d\to
e^+e^-)$, $\mathcal{B}(B^+_u\to \rho^+e^+e^-)$ and
$\mathcal{A}_{FB}(B^+_u\to \rho^+e^+e^-)$ are sensitive to
$\phi_{\spur{R_p}}$ with the first two increasing with
$|\phi_{\spur{R_p}}|$ while the last one decreasing, whereas
$\mathcal{B}(B^+_u\to \pi^+e^+e^-)$ is insensitive to
$\phi_{\spur{R_p}}$.
 From  Figs.\ref{figelplps}(e)-(f), we can find that the
the $\lambda'_{1i3}\lambda'^*_{1i1}$ contributions are significantly
larger than the SM contributions at the all $\hat{s}$ regions  with
theoretical uncertainties included.  While for
$d\mathcal{B}(B^+_u\to \rho^+e^+e^-)/d\hat{s}$, the RPV effect is
significant at the low $\hat{s}$ region. Especially from
Fig.\ref{figelplps}(i)-(j),  we find  the RPV contributions to
$d\mathcal{A}_{FB}^{(')}(B^+_u\to \rho^+e^+e^-)/d\hat{s}$  are quite
different from the SM expectations. All the above observations  can
be extended to Fig.\ref{figulplps}, but  for squark $\tilde{u}_{iL}$
contributions to $B^{0}_{d} \to \mu^{+}\mu^{-}$,  $B^+_u\to \pi^+ \mu^+ \mu^- $
 and  $B^+_u\to \rho^+ \mu^+ \mu^- $,  instead.

The RPV sneutrino exchange contributions to $B^+_u\to\rho^+\ell^+
\ell^-$, $B^+_u\to\pi^+\ell^+ \ell^-$ and $B^0_d\to \ell^+ \ell^-$
are very similar  to each other. We would take
$\lambda^{*}_{i22}\lambda'_{i13}$ contributions as an example, which
is shown by Fig.\ref{figulslp}.  
  Figs.\ref{figulslp}(a)-(b)  show that both 
$\mathcal{B}(B^0_d\to \mu^+\mu^-)$ and $\mathcal{B}(B^+_u\to
\pi^+\mu^+\mu^-)$ are increasing with
$|\lambda^*_{i22}\lambda'_{i13}|$ on  the
$\mathcal{B}$-$|\lambda^*_{i22}\lambda'_{i13}|$ planes,  but
insensitive to $\phi_{\spur{R_p}}$,  and    $\mathcal{B}(B^0_d\to
\mu^+\mu^-)$ can be enhanced to $\sim 1.5\times 10^{-8}$ which
 saturates its present upper limit. 
As shown by Figs.\ref{figulslp}(c) and (d), 
 both $\mathcal{A}_{FB}(B^+_u \to \rho^+ \mu^+ \mu^- )$
and  $\mathcal{B}(B^+_u \to \rho^+ \mu^+ \mu^- )$ are insensitive to
the RPV sneutrino contribution. The scatters of points in
Figs.\ref{figulslp}(c) and (d) are due to theorectical
uncertainties. The similar situation is also found for  the
distributions of $d\mathcal{A}_{FB}(B^+_u \to \rho^+ \mu^+ \mu^-
)/d\hat{s}$ and $d\mathcal{B}(B^+_u \to \rho^+ \mu^+ \mu^-
)/d\hat{s}$. From Figs.\ref{figulslp}(g)-(j), one can find that the present
theoretical uncertainties in our numerical calculation are still
very large,  and the RPV sneutrino exchange contributions are
indistinguishable from the uncertainties. Figs.\ref{figulslp}(e)-(f)
are  the distributions of $d\mathcal{B}^{(\prime)}(B^+_u \to \pi^+
\mu^+ \mu^- )/d\hat{s}$, which  show   the RPV contributions
 distinguishable only in the high $\hat{s}$ region.

As known, to probe new physics effects, it would be very useful to
measure correlative observables, for example, $\mathcal{B}(B^+_u\to
\rho^+\ell^+\ell^-)$, $\mathcal{B}(B^+_u\to \pi^+\ell^+\ell^-)$ and
$\mathcal{B}(B_d\to \ell^+\ell^-)$,  since correlations among these
observables could provide very strict bound on new physics models.

 It should be noted that the present upper limit  of
  $\mathcal{B}(B^+_u\to \pi^+\ell^+\ell^-)<9.1 \times 10^{-8}$ made
  by \textit{BABAR} \cite{Aubert} is based on just $208.9 fb^{-1}$ data,
  which is already just one order higher than the SM expectation.
  Up to this Fall,  Belle and \textit{BABAR} have recorded as much as
  $709 fb^{-1}$ and  $500 fb^{-1}$ data, respectively.
  Searches with these data  will surely improve the upper limits for
  these decays to large extent, which will give much stronger constraints
  on the RPV SUSY  parameter spaces. On the other hand, the
  decays with enhancement of the RPV SUSY might
  be accessible at \textit{BABAR} and Belle,
  especially, the $B^+_u\to \rho^+\ell^+\ell^-$ decay.
For such a measurement, a good $\rho^{+}\to \pi^{+}\pi^{0}$
sub-vertex reconstruction  is necessary to distinguish it from the
much more copious decay $B^{+}\to K^{*+} \ell^+\ell^-  \to
(K^{+}\pi^{0})  \ell^+\ell^-$,  and  may be a challenging task at
\textit{BABAR} and Belle. In the near future at the LHCb,  these
decays might be measured  with  much larger data samples.

\section{Summary}
 We have studied  $B^+_u\to\pi^+\ell^+\ell^-$, $B^+_u\to\rho^+\ell^+\ell^-$
 and $B^{0}_d\to\ell^+\ell^-$ decays extensively in the RPV SUSY.  From the latest experimental
upper limits of $\mathcal{B}(B^+_u\to \pi^+\ell^+\ell^-)$ and
$\mathcal{B}(B^0_d \to \ell^+\ell^-)$, we have obtained the
constrained parameter spaces of the RPV coupling products, and found
these constraints are robust and 
stronger  than the existing ones, which may be useful for further
studies of the RPV SUSY phenomenology. Using the constrained
parameter space, we have given the RPV impacts on the branching
ratios $\mathcal{B}(B^0_d \to \ell^+ \ell^-)$, $\mathcal{B}(B^+_u
\to\pi^+ \ell^+ \ell^- )$ and $\mathcal{B}(B^+_u \to\rho^+ \ell^+
\ell^- )$,  and the forward-backward asymmetries
$\mathcal{A}_{FB}(B^+_u \to \pi^+\ell^+\ell^-)$ and
$\mathcal{A}_{FB}(B^+_u \to\rho^+\ell^+\ell^-)$.

 It is shown that  $\mathcal{B}(B^0_d \to \ell^+\ell^-)$ is very sensitive to
 the RPV  $s$-channel  sneutrino exchange, but not to $t$-channel squark exchange,
 which  could be enhanced to $\sim 10^{-8}$ by the former.
 On the contrary,  $B^+_u\to\rho^+\ell^+\ell^-$ is very sensitive to
 $t$-channel squark exchange, and $ \mathcal{B}(B^+_u\to\rho^+\ell^+\ell^-)$ could be
  enhanced   to $\sim 10^{-6} $,  which could be tested at \textit{BABAR} and Belle.

Furthermore, we have shown the RPV effects on dilepton invariant
mass spectra and the normalized forward-backward asymmetries in
$B^+_u\to\pi^+\ell^+\ell^-$ and $B^+_u\to\rho^+\ell^+\ell^-$ decays.
 In the constrained  parameter space, it is  found that  the RPV couplings $\lambda'_{1i3}\lambda'^*_{1i1}$ and
$\lambda'_{2i3}\lambda'^*_{2i1}$ of squark exchanges could  result in  the distributions of
 $d\mathcal{B}(B^+_u\to\rho^+\ell^+\ell^-)/d\hat{s}$ and  $d\mathcal{A}_{FB}(B^+_u\to\rho^+\ell^+\ell^-)/d\hat{s}$   significantly different from the SM expectations.
 The other four RPV couplings due to the
sneutrino exchanges  give distinguishable contributions to
$d\mathcal{B}(B^+_u\to \pi^+\ell^+\ell^-)/d\hat{s}$ at high
$\hat{s}$ region.  But for the $B^+_u\to\rho^+\ell^+\ell^-$ decays,
the sneutrino exchange contributions to $d\mathcal{B}/d\hat{s}$ and
$d\mathcal{A}_{FB}/d\hat{s}$ are indistinguishable from the
theoretical uncertainties.

With the operation of $B$ factories,  many measurements of rare $B$
decays have put strong  constraints on new physics scenarios beyond
the SM.   With the near future experiments at the LHCb,  one can
access  many more rare decays with high statistics, which will
give more stringent bounds on the products of the RPV couplings.
From measurement of  the correlated decays,  one may eventually
establish indirect evidence for specified new physics model or rule
it out.

The results in this paper could be useful for probing the RPV SUSY
effects and will correlate strongly with searches for the direct RPV
signals at future experiments. Negative experimental evidence for
the deviations from the SM expectations in these decays would result
in strong constraints on these RPV couplings.

\section*{Acknowledgments}
This work is supported by the National Science Foundation under
contract No. 10675039 and the NCET Program sponsored by the Ministry
of Education, China, under No. NCET-04-0656. The work of Ru-Min Wang
is supported by the KRF Grant funded by the Korean Government
(MOEHRD) No. KRF-2005-070-C00030.

\end{document}